\documentstyle[12pt]{article}
\begin{document}
\begin{center}  {\Large\bf Exclusive semileptonic decays of $B$
mesons into\\ light mesons in the relativistic quark model}
 \end{center}
\vskip 5 true mm
\begin{center} { R. N. Faustov, V. O. Galkin and A. Yu. Mishurov\\
\it Russian Academy of Sciences, Scientific Council for Cybernetics,\\
Vavilov Street 40, Moscow 117333, Russia} \end{center}
\vskip 5 true mm

\bigskip
\centerline{\large\bf Abstract}
\smallskip
\noindent  The exclusive semileptonic $B\to \pi(\rho) e\nu$ decays are
studied in the framework of the relativistic quark model based on the
quasipotential approach in quantum field theory. The large recoil
momentum of the final $\pi(\rho)$ meson allows for the expansion of
the decay form factors at $q^2=0$ in the inverse powers of $b$-quark
mass. This considerably simplifies the analysis of these decays. The
$1/m_b$ expansion is carried out up to the first order. It is found
that the $q^2$-dependence of the axial form factor $A_1$ is different
from the $q^2$-behaviour of other form factors. We find $\Gamma(B\to
\pi e\nu)=(3.1\pm 0.6)\times \vert V_{ub}\vert ^2\times
10^{12}$s${}^{-1}$ and $\Gamma(B\to \rho e\nu)=(5.7\pm 1.2)\times
\vert V_{ub}\vert ^2\times 10^{12}$s${}^{-1}$. The relation between
semileptonic and rare radiative $B$-decays is discussed.

\bigskip
\section{Introduction}

The investigation of semileptonic decays of $B$ mesons into light
mesons is important for the determination of the
Cabibbo-Kobayashi-Maskawa matrix element $V_{ub}$, which is the most
poorly studied. At present the value of $V_{ub}$ is mainly determined
from the endpoint of the lepton spectrum in smileptonic $B$-decays
[1]. Unfortunately, the theoretical interpretation of the endpoint
region of the lepton spectrum in inclusive $B\to X_u\ell \bar \nu $
decays is very complicated and suffers from large uncertainties [2].
The other way to determine $V_{ub}$ is to consider exclusive
semileptonic decays $B\to \pi(\rho)e\nu$. These are the
heavy-to-light transitions with a wide kinematic range. In contrast
to the heavy-to-heavy transitions, here we can not expand matrix
elements in the inverse powers of the final quark mass. It is also
necessary to mention that the final meson has a large recoil momentum
almost in the whole kinematical range. Thus the motion of final
$\pi(\rho)$ meson should be treated relativistically. If we consider
the point of maximum recoil of the final meson, we find that
$\pi(\rho)$ bears the large relativistic recoil momentum $\vert{\bf
\Delta}_{max}\vert$ of order of $m_b/2$ and the energy of the same
order. Thus at this kinematical point it is possible to expand the
matrix element of the weak current both in inverse powers of
$b$-quark mass of the initial $B$ meson and in inverse powers of
the recoil momentum $\vert {\bf \Delta}_{max}\vert$ of the final
$\pi(\rho)$ meson. As a result the expansion in powers $1/m_b$ arises
for the $B\to \pi(\rho)$ semileptonic form factor at $q^2=0$, where
$q^2$ is a momentum carried  by the lepton pair. The aim of this
paper is to realize such expansion in the framework of relativistic
quark model. We show that this expansion considerably simplifies the
analysis of exclusive $B\to \pi(\rho)e\nu$ semileptonic decays.

Our relativistic quark model is based on the quasipotential approach
in
quantum field theory with the specific choice of the $q\bar q$
potential.  It provides a consistent scheme for calculation of all
relativistic corrections at a given order of $v^2/c^2$ and allows for
the heavy quark $1/m_Q$ expansion. This model has been applied for the
calculations of meson mass spectra [3], radiative decay widths [4],
pseudoscalar decay constants [5], heavy-to-heavy semileptonic [6] and
nonleptonic [7] decay rates. The heavy quark $1/m_Q$ expansion in our
model for the heavy-to-heavy semileptonic transitions has been
developed in [8] up to $1/m_Q^2$ order. The results are in agreement
with the model independent predictions of the heavy quark
effective theory (HQET) [9]. The $1/m_b$ expansion of rare radiative
decay
form factors of $B$ mesons has been carried out in [10] along the
same lines as in the present paper.

\smallskip

\section{Relativistic quark model}

In the quasipotential approach meson is described by the wave function
of the bound quark-antiquark state, which satisfies the
quasipotential equation [11] of the Schr\"odinger type [12]:
$$\Big(\frac{b^2(M)}{2\mu_{R}}-\frac{{\bf
p}^2}{2\mu_{R}}\Big)\Psi_{M}({\bf p})=\int\frac{d^3 q}{(2\pi)^3}
V({\bf p,q};M)\Psi_{M}({\bf q}),\eqno(1)$$
where the relativistic reduced mass is
$$\mu_{R}=\frac{M^4-(m^2_a-m^2_b)^2}{4M^3};\eqno(2)$$
$$b^2(M)=\frac{[M^2-(m_a+m_b)^2][M^2-(m_a-m_b)^2]}{4M^2},\eqno(3)$$
$m_{a,b}$ are the quark masses; $ M$ is the meson mass; ${\bf p}$ is
the relative momentum of quarks. While constructing the kernel of this
equation $V({\bf p,q};M)$ --- the quasipotential of quark-antiquark
interaction --- we have assumed that effective interaction is the sum
of the one-gluon exchange term with the mixture of long-range
vector and scalar linear confining potentials. We have also assumed
that at large distances quarks acquire universal nonperturbative
anomalous chromomagnetic moments and thus the vector long-range
potential contains the Pauli interaction. The quasipotential is
defined by [3]:
$$V({\bf p,q},M)=\overline{u}_a(p)
\overline{u}_b(-p)\Big\{\frac{4}{3}\alpha_SD_{ \mu\nu}({\bf
k})\gamma_a^{\mu}\gamma_b^{\nu}+V^V_{conf}({\bf k})\Gamma_a^{\mu}
\Gamma_{b;\mu}+V^S_{conf}({\bf k})\Big\}u_a(q)u_b(-q),\eqno(4)$$
where $\alpha_S$ is the QCD coupling constant, $D_{\mu\nu}$ is the
gluon propagator; $\gamma_{\mu}$ and $u(p)$ are the Dirac matrices and
spinors; ${\bf k=p-q}$; the effective long-range vector vertex is
$$\Gamma_{\mu}({\bf k})=\gamma_{\mu}+
\frac{i\kappa}{2m}\sigma_{\mu\nu}k^{\nu},\eqno(5)$$
$\kappa$ is the anomalous chromomagnetic quark moment. Vector and
scalar confining
potentials in the nonrelativistic limit reduce to
$$V^V_{conf}(r)=(1-\varepsilon)(Ar+B),\ \
V^S_{conf}(r)=\varepsilon(Ar+B),\eqno(6)$$
reproducing
$V_{nonrel}^{conf}(r)=V^S_{conf}+V^V_{conf}=Ar+B$, where $\varepsilon$
is the mixing coefficient. The explicit expression for the
quasipotential with the account of the relativistic corrections of
order $v^2/c^2$ can be found in ref. [3].  All the parameters of our
model: quark masses, parameters of linear confining potential $A$ and
$B$, mixing coefficient $\varepsilon$ and anomalous chromomagnetic
quark moment $\kappa$ were fixed from the analysis of meson masses
[3] and radiative decays [4].  Quark masses: $m_b=4.88$ GeV;
$m_c=1.55$ GeV; $m_s=0.50$ GeV; $m_{u,d}=0.33$ GeV and parameters of
linear potential: $A=0.18$ GeV$^2$; $B=-0.30$ GeV have standard values
for quark models.  The value of the mixing coefficient of vector and
scalar confining potentials $\varepsilon=-0.9$ has been primarily
chosen from the consideration of meson radiative decays, which are
very sensitive to the Lorentz-structure of the confining potential:
the resulting leading relativistic corrections coming from vector and
scalar potentials have opposite signs for the radiative Ml-decays
[4]. Universal anomalous chromomagnetic moment of quark $\kappa=-1$
has been fixed from the analysis of the fine splitting of heavy
quarkonia ${ }^3P_J$- states [3].

Recently we have considered the expansion of the matrix elements of
weak
heavy quark currents between pseudoscalar and vector meson states up
to
the second order in inverse powers of the heavy quark masses [8]. It
has been found that the general structure of leading, subleading and
second order $1/m_Q$ corrections in our relativistic model is in
accord with the predictions of HQET. The heavy quark symmetry and QCD
impose rigid constraints on the parameters of the long-range potential
of our model. The analysis of the first order corrections [8] allowed
to fix the value of effective long-range anomalous chromomagnetic
moment of quarks $ \kappa =-1$, which coincides with the result,
obtained from the mass spectra [3]. The mixing parameter of vector
and scalar confining potentials has been found from the comparison of
the second order corrections to be $ \varepsilon =-1$. This value is
very close to the previous one $\varepsilon =-0.9$ determined from
radiative decays of mesons [4]. Therefore, we have got QCD and heavy
quark symmetry motivation for the choice of the main parameters of our
model. The found values of $\varepsilon$ and $\kappa$ imply that
confining quark-antiquark potential has predominantly Lorentz-vector
structure, while the scalar potential is anticonfining and helps to
reproduce the initial nonrelativistic potential.

\smallskip

\section{$B\to \pi(\rho)e\nu $ decay form factors}

The form factors of the semileptonic decays $B\to \pi e\nu$ and $B\to
\rho e\nu$ are defined in the standard way as:
$$\langle\pi(p_\pi)\vert \bar q\gamma_\mu b\vert B(p_B)
\rangle = f_+(q^2)(p_B+p_\pi)_\mu + f_-(q^2)(p_B-p_\pi)_\mu,\eqno(7)$$
$$\langle\rho(p_\rho,e)\vert \bar q\gamma_\mu(1-\gamma^5) b
\vert B(p_B)\rangle = -(M_B+M_\rho)A_1(q^2)e^*_\mu$$
$$+ \frac{A_2(q^2)}{
M_B+M_\rho}(e^* p_B)(p_B+p_\rho)_\mu
+\frac{A_3(q^2)}{M_B+M_\rho}(e^* p_B)(p_B-p_\rho)_\mu +
\frac{2V(q^2)}{M_B+M_\rho}i\epsilon_{\mu\nu \tau\sigma}{e^*}^\nu
p_B^\tau
p_\rho^\sigma,\eqno(8) $$
where $q=p_B-p_{\pi(\rho)}$, $e$ is a polarization vector of $\rho $
meson. In the limit of vanishing lepton mass, the form factors $f_-$
and $A_3$ do not contribute to the decay rates and thus will not be
considered.

The matrix element of the local current $J$ between bound
states in the quasipotential method has the form [13]:
$$\langle \pi (\rho) \vert J_\mu (0) \vert B\rangle =\int \frac{d^3p\,
d^3q}{(2\pi )^6} \bar \Psi_{\pi (\rho)}({\bf p})\Gamma _\mu ({\bf
p},{\bf q})\Psi_B({\bf q}),\eqno(9)$$
where $ \Gamma _\mu ({\bf p},{\bf q})$ is the two-particle vertex
function and  $\Psi_{\pi,B}$ are the meson wave functions projected
onto the positive energy states of quarks.

In the case of semileptonic decays $J_\mu =\bar q \gamma_\mu (1-
\gamma^5) b$ and in order to calculate its matrix element
between meson states it is necessary to consider the contributions to
$\Gamma$ from Figs.~1 and 2.  Thus the vertex functions look like
$$ \Gamma_\mu ^{(1)}({\bf p},{\bf q})=\bar u_q(p_1)\gamma_\mu (1-
\gamma^5)u_b(q_1)(2\pi)^3\delta({\bf p}_2-{\bf q}_2),\eqno(10)$$
and
$$ \Gamma_\mu^{(2)}({\bf p},{\bf
q})=\bar u_q(p_1)\bar u_q(p_2) \Bigl\{\gamma_{1\mu}
(1-\gamma_1^5)\frac{\Lambda_b^{(-)}( k_1)}{\varepsilon
_b(k_1)+\varepsilon_b(p_1)}\gamma_1^0V({\bf p}_2-{\bf q}_2)$$
$$+V({\bf p}_2-{\bf q}_2)\frac{\Lambda_q^{(-)}(k_1')}{
\varepsilon_q(k_1')+ \varepsilon_q(q_1)}\gamma_1^0\gamma_{1\mu}
(1-\gamma_1^5)\Bigr\}u_b(q_1) u_q(q_2),\eqno(11)$$
where ${\bf k}_1={\bf p}_1-{\bf\Delta};\qquad {\bf k}_1'={\bf
q}_1+{\bf\Delta};\qquad {\bf\Delta}={\bf p}_B-{\bf p}_{\pi (\rho)};
\qquad \varepsilon (p)=(m^2+{\bf p}^2)^{1/2}$;
$$\Lambda^{(-)}(p)=\frac{\varepsilon(p)-\bigl( m\gamma
^0+\gamma^0({\bf \gamma p})\bigr)}{ 2\varepsilon (p)}.$$
Note that the contribution $\Gamma^{(2)}$ is the consequence of the
projection onto the positive-energy states. The form of the
relativistic corrections resulting from the vertex function
$\Gamma^{(2)}$ is explicitly dependent on the Lorentz-structure of
$q\bar q$-interaction.

It is convenient to consider the decay $B\to\pi(\rho)e\nu$ in the $B$
meson rest frame. Then the wave function of the final $\pi(\rho)$
meson
moving with the recoil momentum ${\bf\Delta}$ is connected with the
wave function at rest by the transformation [13]
$$\Psi_{\pi(\rho)\,{\bf\Delta}}({\bf
p})=D_q^{1/2}(R_{L{\bf\Delta}}^W)D_q^{1/2}(R_{L{
\bf\Delta}}^W)\Psi_{\pi(\rho)\,{\bf 0}}({\bf p}),\eqno(12)$$
where $D^{1/2}(R)$ is the well-known rotation matrix and $R^W$ is the
Wigner rotation.

The meson wave functions in the rest frame have been calculated by
numerical
solution of the quasipotential equation (1) [14]. However, it is more
convenient to use analytical expressions for meson wave functions. The
examination of numerical results for the ground state wave functions
of mesons
containing at least one light quark has shown that they can be well
approximated
by the Gaussian functions
$$\Psi_M({\bf p})\equiv \Psi_{M\,{\bf 0}}({\bf p})=\Bigl({4\pi\over
\beta_M^2}
\Bigr)^{3/4}\exp\Bigl(-{{\bf p}^2\over 2\beta_M^2}\Bigr),\eqno(13)$$
with the deviation less than 5\%.

The parameters are

$$\beta_B=0.41\ {\rm GeV};\qquad\beta_{\pi(\rho)}=0.31\ {\rm GeV}.$$

Substituting the vertex functions (10), (11), with the account of
the Lorentz transformation of final meson wave function (12), in the
matrix element (9) we can get  the expressions for the semileptonic
decay form factors defined in eqs.~(7) and (8). They are very
cumbersome and
lengthy especially those which arise from the vertex function
$\Gamma^{(2)}$ in Fig.~2, because they explicitly depend on the form
of the $q\bar q$-interaction potential and it is necessary to
integrate both with respect to $d^3p$ and $d^3q$ (see eq.~(9)).
However, as it was already mentioned in the introduction, at the point
of maximum recoil of final $\pi(\rho)$ meson the large value of recoil
momentum $\vert {\bf\Delta}_{max}\vert\sim m_b/2$ allows for the
expansion in powers of $1/m_b$ for the decay form factors $f_+(0)$,
$A_1(0)$, $A_2(0)$ and $V(0)$. This expansion leads to considerable
simplification of the formulae. The large value of recoil momentum
$\vert{\bf\Delta}_{max}\vert$ permits to neglect ${\bf p}^2$  in
comparison
with ${\bf\Delta}_{max}^2$ in the quark energy
$\varepsilon_q(p+\Delta)$ of final meson in the expressions for the
form factors originating from $\Gamma^{(2)}$ \footnote{Such
approximation corresponds to the omitting terms of the third order in
$1/m_b$ expansion.}. Thus we can perform one of the integrations in
the
current matrix element (9) using the quasipotential equation as in the
case of the heavy final meson [4,6]. As a result, we get compact
formulae suitable for numerical treatment. Neglecting terms of the
second order in $1/m_b$ expansion we get for the form factor of $B\to
\pi e\nu$ decay:
$$ f_+(0)=\sqrt{\frac{E_\pi}{ M_B}} \int\frac{d^3p}{
(2\pi)^3} \bar\Psi_\pi\left({\bf p}+\frac{2\varepsilon_q}{
E_\pi+M_\pi}{\bf\Delta}_{max}\right)
\sqrt{\frac{\varepsilon_q(p+\Delta_{max})+m_q}{
2\varepsilon_q(p+\Delta_{max})}} $$  $$\times
\biggl\{1+\frac{M_B-E_\pi}{\varepsilon_q(p+\Delta_{max})+m_q}
+\frac{3}{2}\frac{({\bf p\Delta}_{max})}{{\bf
\Delta}_{max}^2}+ \frac{p_x^2+p_y^2}{(\varepsilon_q(p) +m_q)(
\varepsilon_q(p+\Delta_{max})+m_q)} $$
$$ - 2\left(\frac{p_x^2+p_y^2}{(\varepsilon_q(p)+m_q)^2}
\frac{2}{\varepsilon_b(\Delta_{max})+m_b}- \frac{({\bf
p\Delta}_{max})}
{{\bf\Delta}_{max}^2}\frac{1}{\varepsilon_q(p) +m_q}\left( 1+
\frac{M_B-E_\pi}{\varepsilon_b(\Delta_{max})+m_b} \right)\right)$$
$$ \times\left(M_B+M_\pi-\varepsilon_b(p)-
\varepsilon_q(p)-2\varepsilon_q\left(p+ \frac{2\varepsilon_q}{
E_\pi+M_\pi}\Delta_{max}\right)\right)\biggr\} \Psi_B({\bf
p}),\eqno(14)$$
where we have set [3] the mixing parameter (6) of vector and scalar
confining potentials $\varepsilon=-1$ and long-range anomalous
chromomagnetic quark moment (5) $\kappa=-1$;
$$ \vert {\bf\Delta}_{max}\vert=\frac{M_B^2-M_\pi^2}{2M_B}; \qquad
E_\pi=\frac{M_B^2+M_\pi^2}{2M_B};\eqno(15)$$
and $z$-axis is chosen in the direction of ${\bf\Delta}$.

Similar expressions can be written for the form factors of $B\to \rho
e\nu$ decay at $q^2=0$. They have the same structure as (14) and will
be given elsewhere [15].

Let us proceed further and for the sake of consistensy carry out the
complete expansion of form factor (14) in inverse powers of $b$-quark
mass.

The mass of $B$ meson has the following expansion in $1/m_b$ [9]
$$M_B=m_b+\bar\Lambda+O\left(\frac{1}{ m_b}\right),\eqno(16)$$
In our model parameter $\bar\Lambda $ is equal to the mean value of
the light quark energy inside the $B$ meson
$\bar\Lambda=\langle\varepsilon_q\rangle_B\approx 0.54$~GeV [8].

Now we use the Gaussian approximation for the wave functions (13).
Then shifting the integration variable ${\bf p}$ in (14) by
$-\frac{\varepsilon_q}{ E_\pi+M_\pi}{\bf\Delta}_{max}$, we can factor
out the ${\bf\Delta}$ dependence of the meson wave function
overlap in form factor $f_+$. The result can be written in the form
$$f_+(0)={\cal F}_+({\bf\Delta}_{max}^2)\exp(-\zeta {\bf
\Delta}_{max}^2),\eqno(17)$$
where $\vert{\bf\Delta}_{max}\vert$ is given by (15) and
$$\zeta{\bf\Delta}_{max}^2=\frac{2\tilde\Lambda^2{\bf\Delta}_{max}^2}
{(\beta_B^2+\beta_\pi^2)(E_\pi+M_\pi)^2}=\frac{\tilde\Lambda^2}{
\beta_B^2 }\eta\left(\frac{M_B-M_\pi}{M_B+M_\pi}\right)^2,\eqno(18)$$
here $\eta = \frac{2\beta_B^2}{\beta_B^2+\beta_\pi^2}$
and $\tilde\Lambda$ is equal to the mean value of light quark energy
between $B$ and $\pi$ meson states:
$$\tilde\Lambda =\langle\varepsilon_q\rangle = \frac{1}{\pi}
\frac{m_q^2}{\beta_\pi\sqrt{\eta}}e^zK_1(z)\approx 0.53\ {\rm GeV},
\eqno(19)$$
where $z=\frac{m_q^2}{2\eta \beta_\pi^2}$.

Substituting the Gaussian wave functions (13) in the expression for
the form factor (14) and taking into account (16)--(18), we get up to
the first order in $1/m_b$ expansion:
$$ f_+(0)= N\exp\left(-\frac{\tilde\Lambda^2}{\beta_B^2}
\right) \biggl(1+ \frac{1}{ m_b}\biggl(4\frac{\tilde \Lambda^2}{
\beta_B^2}\eta M_\pi+\frac{2}{3}\left\langle \frac{{\bf p}^2}{\bar
\varepsilon_q +m_q}\right\rangle\left(1+\frac{6}{\sqrt{5}+2}\right)$$
$$-\frac{4}{3(\sqrt{5}+2)}\left\langle\frac{{\bf
p}^2}{(\bar\varepsilon_q
+ m_q)^2}\right\rangle \left(\bar\Lambda +M_\pi+3m_q \right)
+\tilde\Lambda \eta\biggl(2\left(1 +\frac{2}{\sqrt{5}+2} \right)$$
$$ \times\biggl(\left\langle\frac{1}{\bar\varepsilon_q
+m_q}\right\rangle
(\bar\Lambda +M_\pi+3m_q)
-3+\frac{1}{3} \left\langle\frac{{\bf p}^2}{ \bar\varepsilon_q
(\bar\varepsilon_q+m_q)}\right\rangle \biggr) -
\frac{1}{2}\biggr)\biggr)\biggr),\eqno(20)$$
where $N=\left(\frac{2\beta_B\beta_\pi}{\beta_B^2+
\beta_\pi^2}\right)^{3/2}
=\left(\frac{\beta_\pi}{\beta_B}\eta\right)^{3/2}$ is due to the
normalization of Gaussian wave functions in (21); $\bar\varepsilon_q
=\sqrt{{\bf p}^2+m_q^2+\tilde\Lambda^2\eta^2}$, i.~e.
the energy of light quarks in final  meson acquires additional
contribution from the recoil momentum.

\smallskip

\section{Results and discussion}

Using the parameters of Gaussian wave functions (13) in the expression
(20) for the $B\to \pi$ transition form factor $f_+(0)$ and in the
similar expressions [15] for the $B\to \rho$ transition form factors
$A_1(0)$, $A_2(0)$ and $V(0)$ we get
$$\begin{array}{rcl} f_+^{B\to \pi}(0)&=&0.21\pm 0.02\qquad V^{B\to
\rho}(0)
=0.29\pm 0.03\\ A_1^{B\to\rho}(0)&=&0.26\pm 0.03\qquad
A_2^{B\to \rho}(0)=0.30\pm 0.03.\end{array}\eqno(21)$$
The theoretical uncertainty in (21) result mostly from the
approximation of the wave functions by Gaussians (13) and does not
exceed 10\% of form factor values. We have also calculated the second
order terms in $1/m_b$ expansion of form factors [15] and found them
to be small (less than 5\% of form factor values).

We compare our results (21) for the form factors of $B\to
\pi(\rho)e\nu$ decays with the predictions of quark models [16,17] and
QCD sym rules [18,19] in Table 1. There is an agreement between our
value of $f_+^{B\to \pi}(0)$ and QCD sum rule predictions, while our
$B\to \rho e\nu$ form factors are approximately 1.5 times less than
QCD sum rule results.

To calculate the $B\to \pi(\rho)$ semileptonic decay rates it is
necessary to determine the $q^2$-dependence of the form factors.
Analysing the expressions (14), (17), (18), (20) for the form factor
$f_+$ and similar expressions [15] for $A_1$, $A_2$ and $V$, we find
that the $q^2$-dependence of these form factors near $q^2=0$ is given
by
$$f_+(q^2)=\frac{M_B+M_\pi}{2\sqrt{M_B M_\pi}}\tilde\xi(w){\cal F}_+
({\bf\Delta}_{max}^2),\eqno(21)$$
$$A_1(q^2)=\frac{2\sqrt{M_B M_\rho}}{M_B+M_\rho}\frac{1}{2}(1+w)
\tilde\xi(w){\cal A}_1 ({\bf\Delta}_{max}^2),\eqno(22)$$
$$A_2(q^2)=\frac{M_B+M_\rho}{2\sqrt{M_B M_\rho}}\tilde\xi(w){\cal A}_2
({\bf\Delta}_{max}^2),\eqno(23)$$
$$V(q^2)=\frac{M_B+M_\rho}{2\sqrt{M_B M_\rho}}\tilde\xi(w){\cal V}
({\bf\Delta}_{max}^2),\eqno(24)$$
where $w=\frac{M_B^2+M_{\pi(\rho)}^2-q^2}{2M_B M_{\pi(\rho)}}$; ${\cal
F}_+({\bf\Delta}_{max}^2)$ is defined by (17) and ${\cal A}_{1,2}({\bf
\Delta}_{max}^2)$, ${\cal V}({\bf\Delta}_{max}^2)$ are defined
similarly. We have introduced the function
$$\tilde\xi(w)=\left({2\over w+1}\right)^{1/2}\exp
\left(-\eta\frac{\tilde\Lambda^2}{\beta^2}\frac{w-1}{w+1}\right),
\eqno(25)$$
which in the limit of infinitely heavy  quarks in the initial and
final mesons coinsides with the Isgur-Wise function of our model
[8]. In this limit eqs.~(21)--(24) reproduce the leading order
prediction of HQET [9].

It is important to note that the form factor $A_1$ in (22) has a
different $q^2$-dependence than the other form factors (21), (23),
(24). In the quark models it is usually assumed the pole [16] or
exponential [17] $q^2$-behaviour for all form factors. However, the
recent QCD sum rule analysis indicate that the form factor $A_1$ has
$q^2$-dependence different from other form factors [18,19]. In [19] it
even decreases with the increasing $q^2$ as
$$A_1(q^2)\simeq \left(1-\frac{q^2}{M_b^2}\right)A_1(0) \simeq
\frac{2M_B M_\rho}{(M_B+M_\rho)^2}(1+w)A_1(0). \eqno(26)$$
Such behaviour corresponds to replacing $\tilde\xi(w)$ in (22) by
$\tilde\xi(w_{max})$.

We have calculated the decay rates of $B\to \pi(\rho)e\nu$ using our
form factor values at $q^2=0$ and the $q^2$-dependence (21)--(25) in
the whole kinematical region (model A). We have also used the usial
pole dependence for form factors $f_+(q^2)$, $A_2(q^2)$, $V(q^2)$ and
$A_1(q^2)=\frac{2M_B M_\rho}{(M_B+M_\rho)^2}(1+w) \frac{A_1(0)}{
1-q^2/m_P^2}$ (model B), which corresponds to replacing the function
$\tilde\xi(w)$ (25) by the pole form factor. The results are presented
in Table 2 in comparison with the quark model [16,17] and QCD sum rule
[18,19] predictions. We see that our results for the above mentioned
models A and B of form factor $q^2$-dependence coincide within errors.
The ratio of the rates $\Gamma(B\to \rho e\nu)/\Gamma(B\to \pi e\nu)$
is considerably reduced in our model compared to the BSW [16] and ISGW
[17] models, with the simple pole and exponential $q^2$-behaviour of
all form factors. Meanwhile our prediction for this ratio is in
agreement with QCD sum rule results [18,19]. The absolute values of
the rates $\Gamma(B\to\pi e\nu)$ and $\Gamma(B\to\rho e\nu)$ in our
model are close to those from QCD sum rules [19]. The predictions for
the rates with longitudinally and transversely polarized $\rho$ meson
differ considerably in these approaches. This is mainly due to
different $q^2$-behaviour of $A_1$ (see (22), (26) or usual pole
dominance model [16]). Thus the measurement of the ratios $\Gamma(B
\to\rho e\nu)/\Gamma(B\to\pi e\nu)$ and $\Gamma_L/\Gamma_T$ should
provide the test of $q^2$-dependence of $A_1$ and may discriminate
between these approaches.

The differential decay spectra $\frac{1}{\Gamma}\frac{d\Gamma}{d x}$
for $B\to \pi(\rho)$ semileptonic transitions, where
$x=\frac{E_l}{M_B}$
and $E_l$ is lepton energy, are presented in Fig.~3 (see also [23]).

We can use our results for $V$ and $A_1$ to test the HQET relation
[20] between the form factors of the semileptonic and rare radiative
decays of $B$ mesons. Isgur and Wise [20] have shown that in the limit
of infinitely heavy $b$-quark mass an exact relation connects the form
factors $V$ and $A_1$ with the rare radiative decay $B\to\rho\gamma$
form factor $F_1$ defined by:
$$ \langle \rho(p_\rho,e)\vert \bar ui\sigma _{\mu \nu}q^\nu
P_Rb\vert B(p_B)\rangle=i\epsilon_{\mu \nu \tau \sigma }e^{*\nu
}p_B^\tau p_\rho^\sigma F_1(q^2)$$ $$ +\big[e_\mu
^*(M_B^2-M_\rho^2)-(e^*
q)(p_B+p_\rho)_\mu\big]G_2(q^2).\eqno(27)$$
This relation is valid for $q^2$ values sufficiently close to
$q^2_{max}=(M_B-M_\rho)^2$ and reads:
$$F_1(q^2)=\frac{q^2+M_B^2-M_\rho^2}{ 2M_B} \frac{V(q^2)}{
M_B+M_\rho}+\frac{M_B+M_\rho}{2M_B}A_1(q^2).\eqno(28)$$
It has been argued in [21,22,19], that in these processes the soft
contributions dominate over the hard perturbative ones, and thus the
Isgur-Wise relations (28) could be extended to the whole range of
$q^2$. In [10] we developed $1/m_b$ expansion for the rare radiative
decay form factor $F_1(0)$ using the same ideas as in the present
discussion of semileptonic decays. It was shown that Isgur-Wise
relation (28) is satisfied in our model at leading order of $1/m_b$
expansion. The found value of the form factor of rare radiative decay
$B\to\rho\gamma$ up to the second order in $1/m_b$ expansion is [10]
$$F_1^{B\to\rho}(0)=0.26\pm 0.03.\eqno(29)$$
Using (28) and the values of form factors (21) we find
$$F_1^{B\to\rho}(0)=0.27\pm 0.03,\eqno(30)$$
which is in accord with (29). Thus we conclude that $1/m_b$
corrections do not break the Isgur-Wise relation (28) in our model.

\smallskip

\section{Conclusions}

We have investigated the semileptonic decays of $B$ mesons into light
mesons. The recoil momentum of final $\pi(\rho)$ meson is large
compared to the $\pi(\rho)$ mass almost in the whole kinematical
range. This requires the completely relativistic treatment of these
decays. On the other hand, the presence of large recoil momentum,
which for $q^2=0$ is of order of $m_b/2$, allows for the $1/m_b$
expansion of weak decay matrix element at this point. The
contributions to this expantion come both from heavy $b$-quark mass
and large recoil momentum of light final meson.

Using the quasipotential approach in quantum field theory and the
relativistic quark model, we have performed the $1/m_b$ expansion of
the semileptonic decay form factors at $q^2=0$ up to the first order.
We have determined the $q^2$-dependence of the form factors near
$q^2=0$. It has been found that the axial form factor $A_1$ has a
$q^2$-behaviour different from other form factors (see (21)--(24)).
This is in agreement with recent QCD sum rule results [18,19]. The
ratios $\Gamma(B\to\rho e\nu)/\Gamma(B\to\pi e\nu)$ and
$\Gamma_L/\Gamma_T$ are very sensetive to the $q^2$-dependence of
$A_1$, and thus their experimental measurement may discriminate
between different approaches.

We have considered the relation between semileptonic decay form
factors and rare radiative decay form factor [20], obtained in the
limit of infinitely heavy $b$-quark. It has been found that in our
model $1/m_b$ corrections do not break this relation.

\bigskip
\noindent {\bf Aknowledgements}

\medskip

We express our gratitude to B.~A.~Arbuzov, M.~A.~Ivanov,
J.~G.~K\"orner, V.~A.~Matveev, M.~Neubert, V.~I.~Savrin, B.~Stech for
the interest in our work and helpful discussions of the results. This
research was supported in part by the Russian Foundation for
Fundamental Research under Grant No.94-02-03300-a.

\bigskip
\newpage
\noindent {\Large\bf References}

\medskip

\frenchspacing

\begin{enumerate}
\item J. Bartelt et al. (CLEO Collaboration), Phys. Rev. Lett. 71
(1993) 4111
\item T. Mannel and M. Neubert, Phys. Rev. D 50 (1994) 2037; M.
Neubert, Phys. Rev. D 49 (1994) 3392; A.~F. Falk, E. Jenkins, A.~V.
Manohar and M.~B. Wise, Phys. Rev. D 49 (1994) 4553; I. Bigi, M.
Shifman, N. Uraltsev and A. Vainshtein, Preprint CERN-TH.7159/94
(1994)
\item V.~O. Galkin, A.~Yu. Mishurov and R.~N. Faustov, Yad. Fiz. 55
(1992) 2175
\item V.~O. Galkin and R.~N. Faustov, Yad. Fiz. 44 (1986) 1575; V.~O.
Galkin, A.~Yu. Mishurov and R.~N. Faustov, Yad. Fiz. 51 (1990) 1101
\item V.~O. Galkin, A.~Yu. Mishurov and R.~N. Faustov, Yad. Fiz. 53
(1991) 1676;
\item V.~O. Galkin, A.~Yu. Mishurov and R.~N. Faustov, Yad.
Fiz. 55 (1992) 1080
\item R.~N. Faustov, V.~O. Galkin and A.~Yu. Mishurov, in: Proc.
Seventh Int. Seminar "Quarks'92", eds. D.~Yu. Grigoriev et al.
(World Scientific, Singapore 1993), p.~326
\item R.~N. Faustov and V.~O. Galkin, Z. Phys. C in press
\item M. Neubert, Phys. Rep. 245 (1994) 259
\item R.~N. Faustov and V.~O. Galkin, Preprint JINR E2-94-438, Dubna
(1994)
\item A.~A. Logunov and A.~N. Tavkhelidze, Nuovo Cimento 29 (1963) 380
\item A.~P. Martynenko and R.~N. Faustov, Teor. Mat. Fiz. 64 (1985)
179
\item R.~N. Faustov, Ann. Phys. 78 (1973) 176; Nuovo Cimento 69
(1970) 37
\item V.~O. Galkin and R.~N. Faustov, Teor. Mat. Fiz. 85 (1990) 155
\item R.~N. Faustov, V.~O. Galkin and A.~Yu. Mishurov, in preparation
\item M. Wirbel, B. Stech and M. Bauer, Z. Phys. C 29 (1985) 637
\item N. Isgur, D. Scora, B. Grinstein and M.~B. Wise, Phys. Rev. D 39
(1989) 799
\item P. Ball, Phys. Rev. D 48 (1993) 3190
\item S. Narison, Preprint CERN-TH.7237/94 (1994)
\item N. Isgur and M.~B. Wise, Phys. Rev. D 42 (1990) 2388
\item G. Burdman and J.~F. Donoghue, Phys. Lett. B 270 (1991) 55
\item S. Narison, Preprint CERN-TH.7166/94 (1994)
\item R.~N. Faustov, V.~O. Galkin and A.~Yu. Mishurov, in preparation
\end{enumerate}

\nonfrenchspacing
\newpage
\noindent {\bf Table~1} Semileptonic $B\to \pi$ and $B\to \rho$ decay
form factors.

\bigskip
\begin{tabular}{ccccc}
\hline
Ref. & $f_+^{B\to\pi}(0)$ & $A_1^{B\to\rho}(0)$ & $A_2^{B\to\rho}(0)$
& $V^{B\to\rho}(0)$ \\
\hline
our results & $0.21\pm 0.02$ & $0.26\pm 0.03$ & $0.30\pm 0.03$ &
$0.29\pm 0.03$ \\
16 & 0.33 & 0.28 & 0.28 & 0.33 \\
17 & 0.09 & 0.05 & 0.02 & 0.27 \\
18 & $0.26\pm 0.02$ & $0.5\pm 0.1$ & $0.4\pm 0.2$ & $0.6\pm 0.2$ \\
19 & $0.23\pm 0.02$ & $0.38\pm 0.04$ & $0.45\pm 0.05$ & $0.45\pm
0.05$ \\
\hline
\end{tabular}

\bigskip

\medskip

\bigskip

\noindent {\bf Table~2} Semileptonic decay rates $\Gamma(B\to\pi
e\nu)$, $\Gamma(B\to \rho e\nu)$  ($\times \vert
V_{ub}\vert^2\times 10^{12}$s${}^{-1}$) and the ratio of the rates for
longitudinally and transversely polarized $\rho$ meson.

\bigskip

\begin{tabular}{cccc}
\hline
Ref. & $\Gamma (B\to\pi e\nu)$ & $\Gamma(B\to\rho e\nu)$ &
$\Gamma_L/\Gamma_T$ \\
\hline
our results &   &   &   \\
model A & $3.1\pm 0.6$ & $5.7\pm 1.2$ & $0.6\pm 0.3$ \\
model B & $3.0\pm 0.6$ & $5.2\pm 1.2$ & $0.6\pm 0.3$ \\
16 & 7.4 & 26 & 1.34 \\
17 & 2.1 & 8.3 & 0.75 \\
18 & $5.1\pm 1.1$ & $12\pm 4$ & $0.06\pm 0.02$ \\
19 & $3.6\pm 0.6$ & $5.1\pm 1.0$ & $0.13\pm 0.08$ \\
\hline
\end{tabular}

\bigskip

\bigskip

\bigskip

\noindent {\bf Figure captions}

\smallskip
\noindent {\bf Fig.~1} Lowest order vertex function

\noindent {\bf Fig.~2} Vertex function with the account of the quark
interaction. Dashed line corresponds to the effective potential (4).
Bold line denotes the negative-energy part of the quark propagator.

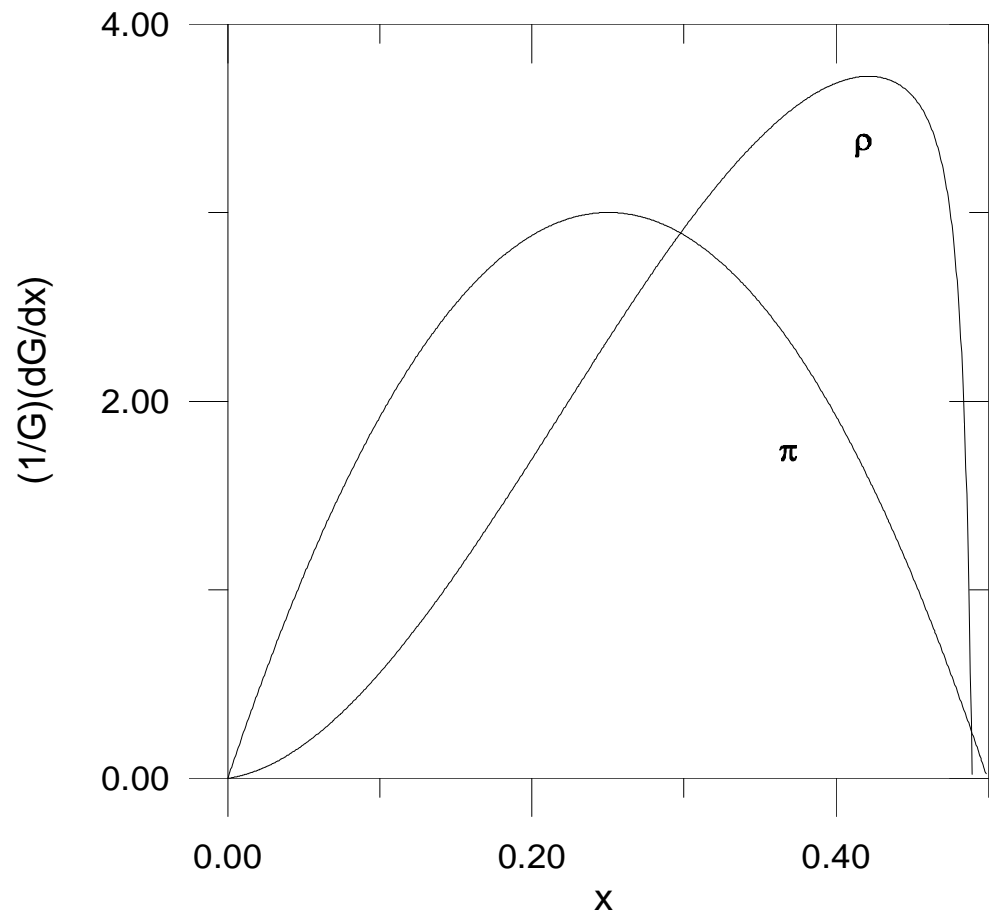
\noindent {\bf Fig.~3} The differential decay spectra (model A)
$\frac{1}{\Gamma}\frac{d\Gamma}{d x}$
for $B\to \pi(\rho)$ semileptonic transitions, where
$x=\frac{E_l}{M_B}$
and $E_l$ is lepton energy. Absolute rates $\frac{d\Gamma}{d x}$
can be obtained  using $\Gamma(B\to\pi(\rho) e\nu)$ from Table~2.

\end{document}